# Albert Einstein and the Marangoni family

Christian Bracco - SYRTE, Observatoire de Paris, Université PSL, CNRS, Sorbonne Université, LNE, 61 avenue de l'Observatoire 75014 Paris - E-mail: Chritian.Bracco@obspm.fr

*Abstract*: The sixteen years old Albert Einstein made friend with Ernestina Marangoni in the summer of 1895 in Pavia. We discuss in this article the unacknowledged link between Albert Einstein and the physicist and professor Carlo Marangoni, Ernestina's uncle, a specialist of capillarity effects.

**1. Ernestina's family**

Ernestina Marangoni (1876-1972) is friend with Albert Einstein and his younger sister Maja. The relation between Ernestina and Maja lasts all their life long and develops in Italy, where Maja and her husband Paul Winteler live from 1922 to 1938 (in Colonnata near Florence). The meeting between Ernestina and Albert took place in the summer of 1895 (or 1896) on the banks of the Ticino river. [1] The Pavia University Museum holds three letters from Albert to Ernestina, dated 16th August 1946, 7th October 1947 and 1st October 1952. They are reproduced in Fregonese (2005). These letters testify of their friendship and of the pleasant memories that Albert Einstein kept of his stays in Pavia, where his father and his uncle were managing the electrotechnical factory *Einstein, Garrone & C* until 1896 (Sanesi 1982, Biscossa 2005), and in Casteggio with the Marangoni family, where he was spending some holidays. In 1946, he remembered "I mesi felici del mio soggiorno in Italia sono le più belle ricordanze"; and in 1952 "Che bel ricordo è Casteggio".

---

[1] Bernini (1994) mentions the year 1896 for the meeting, from Ernestina's 1949 notes. The year 1895 is mentioned in Fregonese (2005) and seems more probable (Bracco 2017).

Ernestina is born in Casteggio, a little town Southeast and in the vicinity of Pavia, on 29th November 1879. Her father is Giovanni Battista <u>Giulio</u> Luigi Marangoni, born in Pavia on 28th May 1843; [2] he died in Casteggio on 16th October 1918. [3] Child, he is already known by the surname Giulio, as established in the inhabitant's register of the basilica San Michele Maggiore in Pavia where he had been christened. [4] Following Bernini (1994), Giulio was an expert in silkworms. [5] Albert considered him in great respect and he compared him to a « *Leonardo da Vinci* » in his letters to Ernestina. Giulio married Rachel Venco, the daughter of a family of pharmacists in Casteggio, on 11th June 1872. After her studies at the technical Institute Bordoni, Ernestina graduated as a doctor in chemistry at the Pavia University in 1902. She married Edmondo Pelizza in 1903.

Musical afternoons gathered the two friends, Albert playing the violin when Ernestina was playing the pianoforte in Casteggio or in the music saloon of the villa in Pavia (Fregonese 2005). [6] When the young Einstein visits the Marangoni family, from 1895 to 1900, he stays in a little house adjoining their splendid villa in Casteggio, built on the "pistornile" at the beginning of the 19th century in the style of a Venetian palace. He remains there during the traditional celebration of the grape harvest, in September. The link between Albert Einstein and the Marangoni family not only gives evidence of the social and intellectual life of the Einstein family in Pavia, but it unveils a link unacknowledged by now: the link between Albert Einstein and the physicist Carlo Marangoni, Ernestina's uncle (Bracco 2015, 2017).

**2. Carlo Marangoni, physicist and professor.**

Luigi <u>Carlo</u> Giuseppe Marangoni (1840-1925), Giulio's eldest brother, was born in Pavia on 29th April 1840. In his childhood, he used to live in Pavia at 853 via San Michele, with Giulio, their two eldest sisters Amalia and Luigia and their parents Matteo Marangoni, an accountant, and Ernesta Robecchi.

Carlo completes his third year of study in pure Mathematics at the Pavia University whose rector was Giovanni Cantoni, one of the major physicists of the time in Italy. He obtains his diploma on 18th December 1863. In his dissertation, entitled *Sull'ascesa della linfa nelle piante*, Marangoni confronted the theories of Jamin and Dutrochet on the ascent of sap in relation with endosmose phenomena. [7]

Then, he works in thesis with Cantoni, his supervisor, on *Sull'espansione delle goccie d'un liquido galleggianti sulla superficie di altro liquido*. He obtains his doctorate in 1865. His manuscript begins with the famous French proverb: « *Ne*

---

[2] Baptismal register of the basilica San Michele Maggiore, Pavia.
[3] Municipal register in Casteggio.
[4] *Stato d'anime*, basilica San Michele Maggiore.
[5] Many details concerning the Marangoni family can be found in Bernini's little book.
[6] Einstein's family lived in a villa at 11 via Ugo Foscolo, the *Cornazzani* palace, where the famous poet Ugo Foscolo used to live in 1808. Luigi Doglia bought the villa in 1873. His suns Roberto and Goffredo inherited it in 1878 (State archives in Pavia).
[7] Folder Carlo Marangoni, archives of the Pavia University.

*cherchez pas midi à quatorze heures* ». By the careful observation of a droplet of essential oil spreading on the surface of water, he refutes prior explanations calling for effects analogous to gravitation, gas expansion, electric action or even unknown forces. He establishes that the phenomenon is only related to surface tension (already known to account for capillarity phenomena), namely the difference in surface tension of the liquids in presence. He details the many experiences he has lead on the great basin of the Tuileries in Paris to support this view.

Carlo becomes in 1865 Cantoni's assistant and publishes Cantoni's 1865-66 lectures (Marangoni 1866). Then he becomes Carlo Matteucci's assistant, the famous professor at the Museum of physics in Florence, which is one of the three components, together with medicine and philology, of the technical Institute of the town. In 1870 he becomes a professor of physics at the *Liceo Dante* until he retires in 1916. During his period of activity, he leads his researches at the technical Institute and publishes about eighty papers in physics as well as several books. A detail account of his researches can be found in (Borghi 2008).

Following a private communication with Fabrizio Bernini (1949-2017), Carlo met his brother Giulio once a year in Casteggio, during the grape harvest:

> … Ernestina un giorno si soffermò sullo zio Carlo Marangoni, narrandomi una vicenda che direttamente lo interessava ....
>
> L'ingegner Carlo Marangoni, dalla Toscana, veniva a soggiornare in Casteggio presso il fratello una volta all'anno in tempo di vendemmia, rito atavico a festoso per l'Oltrepò che affascinava Einstein. Puntualmente dunque, ogni anno, anche tra il 1896 ed il 1900 circa, stringendo amicizia con il giovane futuro scienziato.
>
> Mentre Einstein dormiva in villa Carlo Marangoni, essendo la stessa troppo affollata in quel periodo, pernottava con la moglie ed il figlio presso il cascinale « Fontanone », posto tra i vigneti, quasi sperduto. Era una proprietà dell'ingegner Pasquale Pelizza, il futuro suocero di Ernestina (extract of Bernini's letter, 17th March 2016).

During familial meetings, Carlo must have found much interest in discussing with the young Albert Einstein, who wanted to enter the federal polytechnic school in Zurich (ETHZ) to study physics - and who studied there until 1900. For his part, Carlo being a teacher, he was naturally inclined to pass on his knowledge: he had already written a textbook on experimental physics intended for teachers (Marangoni 1879), which was one of his well-known books.[8] Let us examine a few possible consequences of these meetings on Einstein's early centres of interest.

**3. From capillarity to Brownian motion**

---

[8] Marangoni Obituary published in *Il Nuovo Cimento* n. 2 (1925).

*3.1. Capillarity effects.*

The name of Carlo Marangoni has remained attached in physics to the Gibbs-Marangoni effect, in line with his thesis results. This effect describes a flow of matter appearing when two liquids of different surface tension enter in contact. "Tears of vine" are a consequence of it: in the region of the meniscus, near the wall of the glass, vine raises by capillarity. In that region, alcohol evaporates easily leading to a gradient in surface tension. The liquid moves then towards regions of greater surface tension (surface tension of alcohol being less than water) before falling under gravity. It's tempting to imagine Carlo explaining tears of vine to the young Albert during grape harvest in Casteggio!

Such discussions may have also driven Albert's attention to Sir William Thomson (Lord Kelvin) *Popular conferences and addresses*, a collection of three books that can be found in their original printed edition in Einstein's personal library. The first volume is published in 1889 and is entitled "Constitution of Matter" (Thomson 1889). It begins with a 73 pages chapter dedicated to capillary forces. Thomson writes that his own brother James was the first to give an explanation of tears of vine in 1855. He even reproduces in an Appendix James' communication to the British Association at the Glasgow meeting entitled "On certain curious motions observable on the surfaces of wine and other alcoholic liquors". Carlo may well have given to the young Albert his own point of view concerning this priority dispute raised by Thomson.

Let us note that Carlo Marangoni also dealt with capillarity phenomena in his pedagogical book of 1879 already mentioned. He gives a table of molecular distances deduced from capillarity effects by many authors and attributes the best value of 1/250000 mm to the French physicist Jules Violle. The latter was known to Einstein, who had read his textbook on general physics in 1895, a book bought in the library Hoepli in Milan. Let us emphasize that molecular dimensions have been a major subject of reflexion for Einstein: he first began by discarding their role in his study of molecular forces until 1902, before explaining how to measure them in his thesis in 1905.

Let us finally note that in his 1907 article on capillarity in the Sommerfeld's *Encyclopaedia*, Minkowski mentions Carlo Marangoni at three times, referring to him as a specialist of the subject (Minkowski 1907). Minkowski also gave a lecture on capillary forces at the ETHZ, to which Einstein attended with a great pleasure, following the memories of his classmate Louis Kollros. Maybe these lectures allowed Einstein understanding better what Marangoni already explained to him in a less formal way?

In December 1900, Albert submits to the *Annalen der Physik* his first paper, published in March 1901, entitled "Conclusions drawn from the phenomena of capillarity" (Einstein 1901). Aged 21, the young man had begun without any appointment, at the end of 1900, a thesis on molecular forces. Einstein makes in his article an original (although wrong) assumption: any chemical element could be characterized by a specific constant $c_\alpha$, which plays with respect to molecular forces a role analogous to mass for gravitation. By combining two energetic methods, one issued from mechanics (Lagrange, Gauss) and the other issued from thermodynamics (Thomson,

Gibbs), he obtains the $c_\alpha$ for the elements H, C, O, Cl, Br, I, by using data of about 40 organic molecules and by using the least mean square method (Bracco 2017 and references therein). He extends the subject of his thesis to molecular forces in gases in April 1901 (discarding the role of molecular dimensions), then he submits his dissertation to his supervisor Alfred Kleiner in November or early December 1901 and he finally abandons it in the beginning of 1902 (Bracco, Provost 2018a, 2018b).

*3.2. Brownian motion.*

In 1913, Icilio Guareschi devotes an article on the history of the Brownian motion in the first issue of the journal *Isis*. He writes:

> *Un lavoro più importante sul movimento browniano si deve a Giovanni Cantoni; egli trovò la vera causa di questo movimento che sono precisamente quelle, o molto analoghe a quelle, emesse più recente. Il professore Giovanni Cantoni, il quale per molti anni fu professore di fisica sperimentale nella Università di Pavia, già sino dal 1867 aveva chiaramente ammesso che il movimento browniano dipende dal movimento delle molecole del liquido in cui si trova sospesa la sostanza*" (Guareschi 1913).

Cantoni had published in 1867 his article on Brownian motion in *Il Nuovo Cimento*. He concluded his study by:

> *Ora tutti gli esposti particolari concorrono alla deduzione, che la condizione fisica del moto browniano stia nella diversa velocità che hanno le molecole dei corpi differenti sotto una stessa temperatura. E di tal modo il moto browniano, cosi dichiarato, ci fornisce una delle più belle e dirette dimostrazioni sperimentali dei fondamentali principii della teoria meccanica del calore, manifestando quell'assiduo stato vibratorio che esser deve e nei liquidi e nei solidi ancor quando non si muta in essi la temperatura* (Cantoni 1867).

Note that the journal *Il nuovo Cimento* founded in Pisa, was at that time directed by the professors Matteucci and Piria, but also "e continuato dai professori di scienze fisiche e naturali di Pisa e del Museo di Firenze" as stamped on the front page of the volume. Among them, there must have been Carlo Marangoni, since he was Matteucci's assistant at the Florence Museum. Marangoni would not have missed Cantoni's paper on Brownian motion, a topic that in addition he probably discussed with him a few months before when he was his assistant. Therefore, Albert Einstein could well have been acquainted before 1900 with the kinetic explanation of the Brownian motion through discussions with Marangoni. In May 1905, after having completed his thesis, the explanation came to him suddenly as Einstein later remembered (Shankland 1952). [9] Perhaps previous discussions were also resurfacing?

---

[9] Cantoni relied on the Dulong-Petit and Graham laws relatives to specific heat, which allowed him deducing that the kinetic molecular energy $mv^2/2$ only depends on temperature; then he applied this result to Brownian particles. In 1905, Einstein computes the mean free path of these particles as a function of their dimension, using results of his thesis concerning their diffusion coefficient (discussions with J.-P. Provost).

To conclude, let us remember that Ernestina Marangoni also said that Einstein had given her in 1896 an article entitled *On electricity and the electric current* that he asked her to return « *because it was wrong* » (Fregonese 2005, Bevilacqua, Bordoni 2005). Did he care about critics from Carlo?

**Bibliografia**